\pdfoutput=1

\documentclass{prepr}
\usepackage[backend=bibtex,style=numeric-comp]{biblatex}
\newcommand{\pta}{\ensuremath{p_\mathrm{T}}}
\newcommand{\nev}{\ensuremath{N_{\mathrm{ev}}}}
\newcommand{\nch}{\ensuremath{n_{\mathrm{ch}}}}
\newcommand{\Nch}{\ensuremath{N_{\mathrm{ch}}}}

\newcommand{\ptlead}{\ensuremath{p_\mathrm{T}^\mathrm{lead}}}

\newcommand{\dndpt}{\ensuremath{\frac{1}{\nev}\cdot \frac{1}{2 \pi  p_\mathrm{T}} \cdot \frac{\mathrm{d}^2 \Nch}{\mathrm{d} \eta \mathrm{d} p_\mathrm{T}}}}
\newcommand{\dndeta}{\ensuremath{\frac{1}{\nev} \cdot  \frac{\mathrm{d} \Nch}{\mathrm{d} \eta}}}

\newcommand{\sumpt}{\ensuremath{\langle \sum \pta / \delta\eta \delta\phi \rangle }}
\newcommand{\sumnch}{\ensuremath{\langle \Nch / \delta\eta \delta\phi \rangle }}

\newcommand{\nselbs}{\ensuremath{n_{\mathrm{sel}}^{\mathrm{BL}}}}

\newcommand{\dzerobs}{\ensuremath{d_\mathrm{0}^{\mathrm{BL}}}}

\newcommand{\py}{\textsc{pythia 8}}
\newcommand{\epos}{\textsc{epos}}
\newcommand{\qgsjet}{\textsc{qgsjet-ii}}
\newcommand{\hpp}{\textsc{herwig++}}

\begin{document}
\vspace*{6cm}

\title{ Measurements of minimum bias events,
underlying event and particle production
properties in ATLAS\footnote{Proceedings of the 23rd Low-x Meeting, Sandomierz, Poland, September 1-5, 2015}}

\author{Krzysztof W. Wo\'{z}niak on behalf of the ATLAS Collaboration }
\address{ Institute of Nuclear Physics, Polish Academy of Sciences, ul. Radzikowskiego 152, 31-342 Krak\'{o}w, Poland} 

\maketitle

\vspace*{2cm}

\abstracts{%
  The measurements of the minimum bias events provide valuable information
on the basic properties of the $pp$ interactions. The results
at the new highest energy of $pp$ collisions, \mbox{$\sqrt{s}=13$ TeV,}
obtained using the ATLAS detector, are shown.
They include distributions of charged particle pseudorapidity
density, transverse momentum and multiplicity.
The properties of the underlying event, determined with respect to
a leading high-$\pta$ particle, are also presented.
In both cases the new results are compared with those from earlier studies
of the $pp$ collisions at  \mbox{$\sqrt{s}=7$ TeV.}
}

\vspace*{2cm}

%-------------------------------------------------------------------------------
\section{Introduction}
\label{sec:intro}
%-------------------------------------------------------------------------------

The first measurements of $pp$ collisions at $\sqrt{s}=13$ TeV
by the ATLAS detector~\cite{PERF-2007-01} at the Large Hadron Collider
~(LHC)~\cite{Evans:2008zzb} were performed 
at low instantaneous luminosity
which ensured that the mean number of $pp$ interactions
in the registered events, $\langle \mu \rangle$, was only about 0.005. Such sample of events
is suitable to study the strong interactions in the soft, non-perturbative
QCD region. General properties of the minimum bias events~\cite{ATLAS-CONF-2015-028}
and the activity accompanying a hard scattering 
(the underlying event - UE)~\cite{ATL-PHYS-PUB-2015-019} can be 
compared to predictions of Monte Carlo event generators (MC), which
are later used for simulations of the effects of multiple proton–proton 
interactions. Results of such comparisons can be used to tune free 
parameters of the models leading to a better description of the data.
These MC generators are then extensively used also in the analysis
of processes, which require data collected at high instantaneous luminosity.
In the studies mentioned above several models are compared with the data: 
\py~\cite{Sjostrand:2007gs}, \hpp~\cite{Bahr:2008pv},
\epos~\cite{Porteboeuf:2010um,Werner:2011um} and \qgsjet~\cite{Qgsjet}.
As it is described in more detail in \cite{ATLAS-CONF-2015-028,ATL-PHYS-PUB-2015-019}
for each model its parameters were optimized using the existing 7~TeV data
and the following tunes were used: \cite{ATL-PHYS-PUB-2011-014,Skands:2014pea,
Gieseke:2012ft,Pierog:2013ria,Sherstnev:2008dm,Ball:2012cx,ATL-PHYS-PUB-2014-021,Pumplin:2002vw}.

\begin{figure}
        \begin{center}
       % \subfigure[]{
                \includegraphics[width=0.49\textwidth]{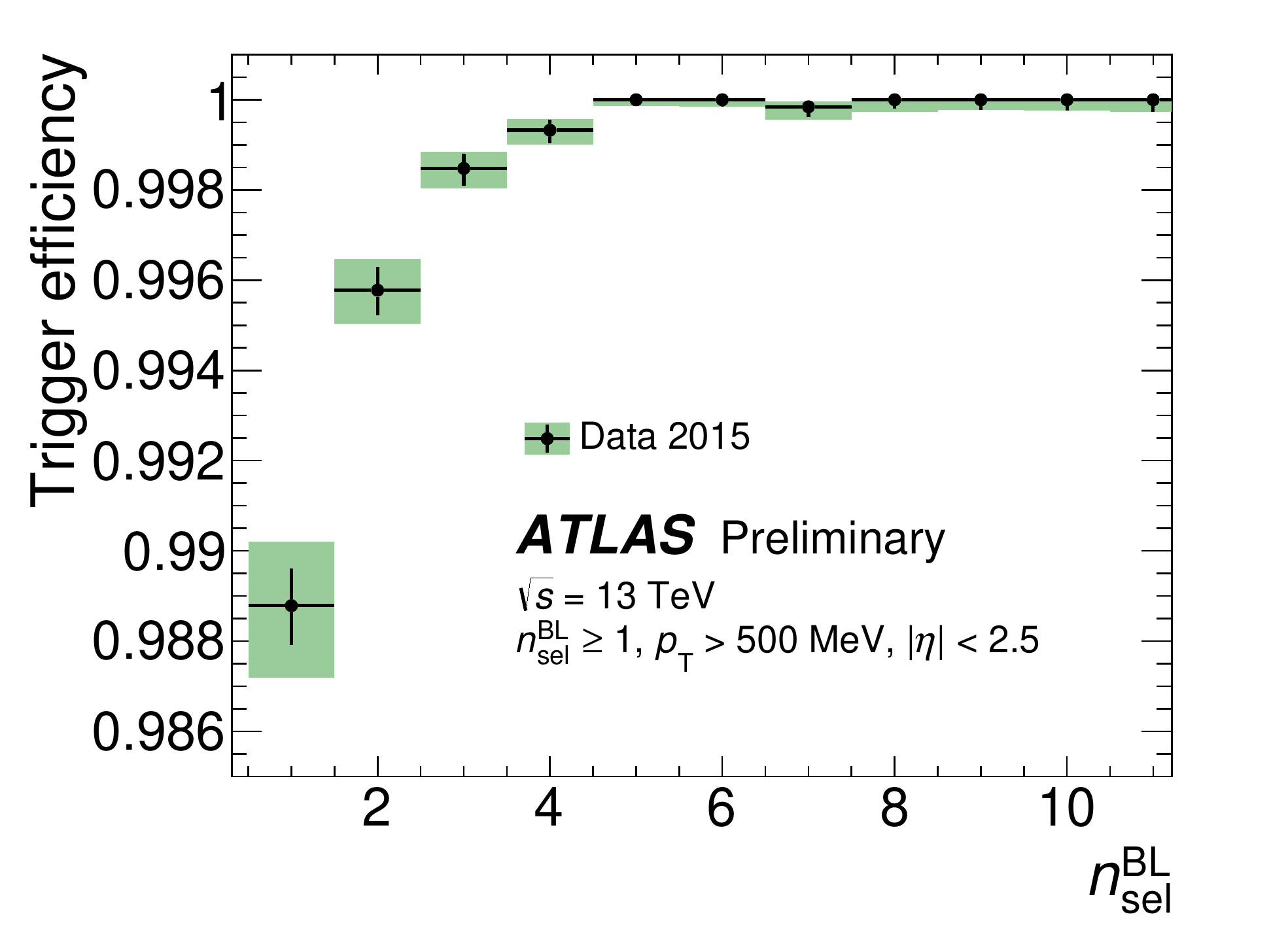}
%                \label{fig:eff-trigger}
       % }
       % \subfigure[]{
                \includegraphics[width=0.49\textwidth]{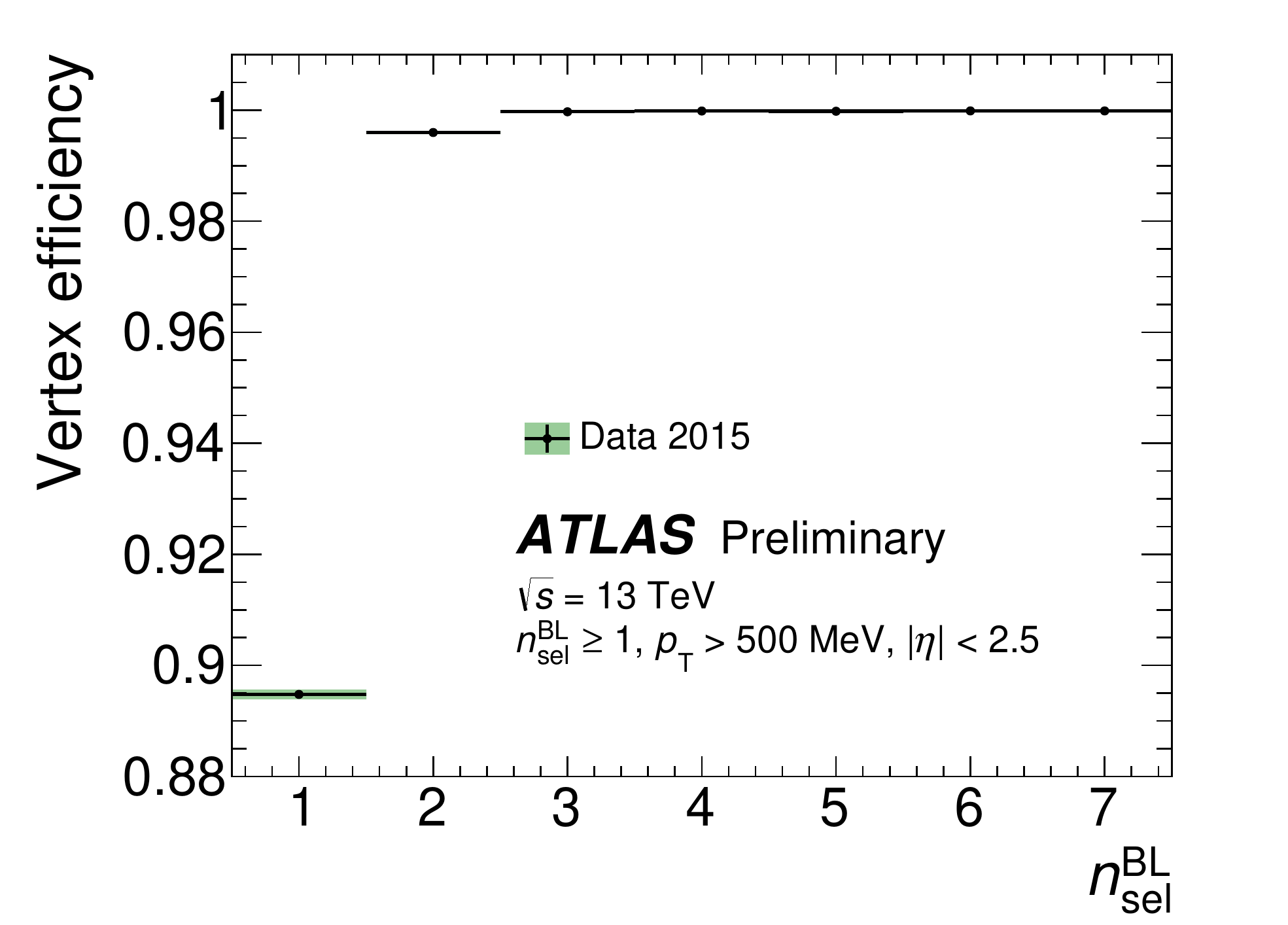}
%                \label{fig:eff-vertex}
       % }
        \end{center}
        \caption[]{(left) Trigger efficiency and (right) vertex reconstruction efficiency 
with respect to the event selection~\cite{ATLAS-CONF-2015-028}, 
as a function of the number of reconstructed tracks 
before the vertex requirement, (\nselbs).
The statistical errors are shown as black lines, the total errors as green shaded areas.}
  \label{fig:eff}
\end{figure}

%-------------------------------------------------------------------------------
\section{ATLAS detector and data selection}
\label{sec:detector}
%-------------------------------------------------------------------------------

The ATLAS detector~\cite{PERF-2007-01}  at the Large Hadron 
Collider (LHC)~\cite{Evans:2008zzb}
is a general purpose apparatus enabling measurements in an almost 
full solid angle, detecting charged particles and the neutral particles
interacting strongly or electromagnetically.
The inner detector (ID) placed in a 2 Tesla magnetic field 
covers the full azimuthal angle and the pseudorapidity
range $|\eta|<2.5$. The barrel part of ID, surrounding the beam pipe
closest to the nominal collision point, consists of 4 layers of 
silicon pixel sensors followed by eight layers of microstrip silicon 
sensors (SCT). It is worth to mention that since 2015 an additional
innermost layer of silicon sensors (Insertable B-Layer - IBL),
installed at the radius of only 33 mm from the beam line, is used.
The size of the pixels in this new layer is smaller than in the next
three layers of the pixel detector.
The barrel part of pixel and SCT detectors is complemented 
at both ends by disks
containing at each side 3 pixel and 9 microstrip sensor layers.
At a larger radial distance from the beam line, behind SCT, 
the tracks of particles are measured in the Transition Radiation 
Tracker (TRT), the last element of the inner detector. 
Outside the inner detector are placed calorimeters and muon detectors,
supplying additional information on particles produced in the collisions.
 
The trigger used to collect the minimum bias data is 
based on the signals from the Minimum Bias Trigger Scintillators (MBTS).
This detector consists of two disks of plastic scintillators divided
into 12 modules each and registers charged particles emitted in 
the range $2.07 < |\eta| < 3.86$. The trigger requires at least one module
with the signal above the threshold in any of the disks.
The efficiency of this trigger, shown in Fig.~\ref{fig:eff}~(left), 
is studied using a fully unbiased control 
trigger requiring only one reconstructed track with $\pta > 200$~MeV.
The efficiency is almost 99\% already for $\nselbs=1$ and very fast 
increases to 100\%.

In order to remove background contamination and reject events with more 
than one $pp$ interaction in the same beam crossing a single primary
vertex with at least two tracks is required. However, in the events with more 
than one reconstructed vertex only those in which the second vertex contains
at least four tracks are rejected. The second vertex with less than four 
tracks is allowed as it is usually due to secondary interactions or
results from splitting of the primary vertex. The vertex reconstruction 
efficiency defined as a fraction of triggered events containing the vertex
is shown in Fig.~\ref{fig:eff}~(right).
Only in the events with $\nselbs=1$ this efficiency is slightly below 90\%, for 
$\nselbs=2$ it reaches 99.7\% and is 100\% for higher multiplicities.

The same data sample is used in both minimum bias and UE studies and it corresponds 
to an integrated luminosity of 170 $\mu\mathrm{b}^{-1}$. For the charged
particle multiplicity analysis at least one track with  $\pta>0.5$~GeV 
and $|\eta|<2.5$ is also required and finally 8,870,790 events are selected. 
This number is reduced in the underlying event study, as the presence of 
a leading track with $\pta>1$~GeV or  $\pta>5$~GeV is in addition requested.

\begin{figure}
        \begin{center}
       % \subfigure[]{
                \includegraphics[width=0.49\textwidth]{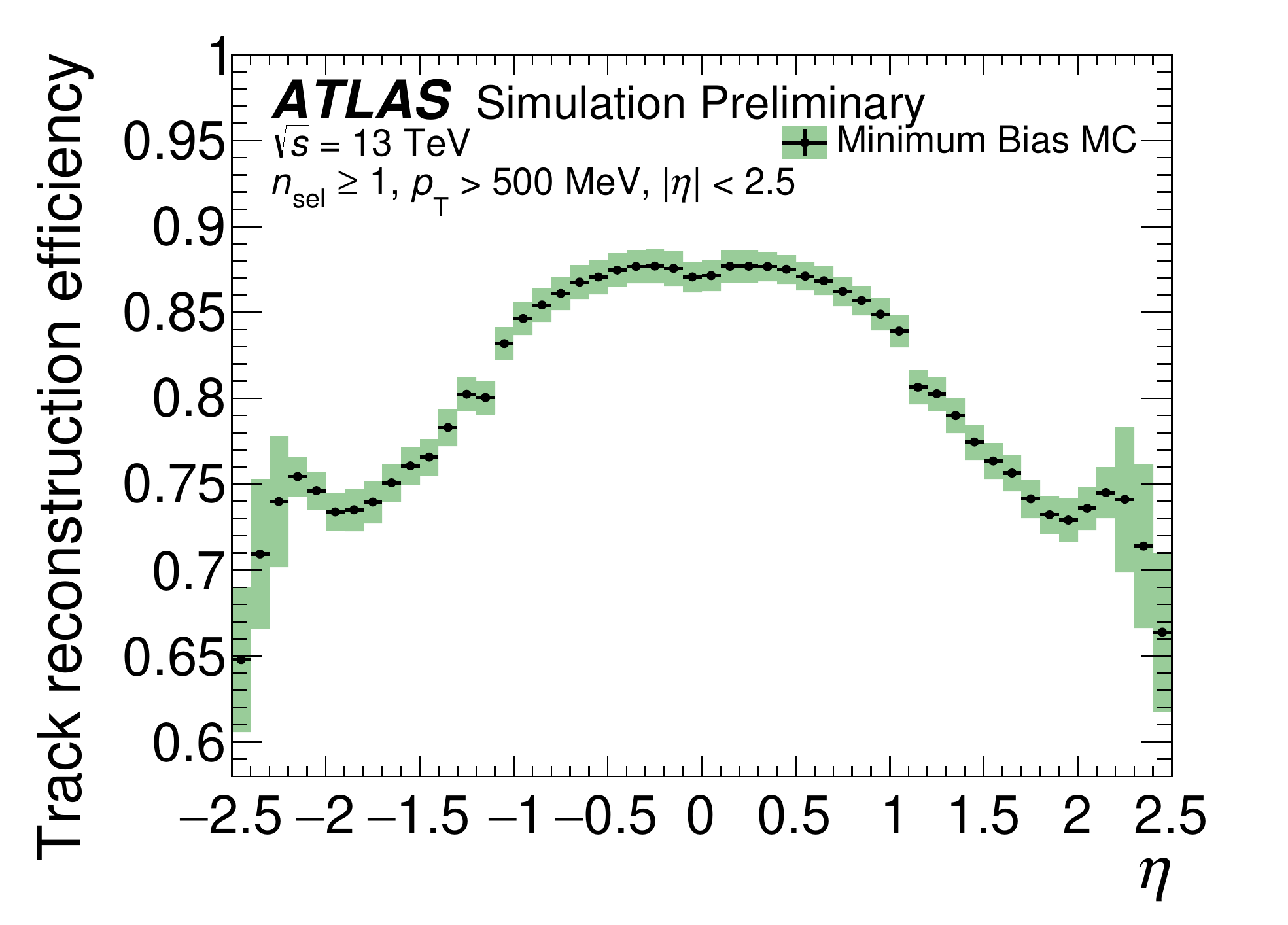}
   %             \label{fig:eff-tracketa}
       % }
       % \subfigure[]{
                \includegraphics[width=0.49\textwidth]{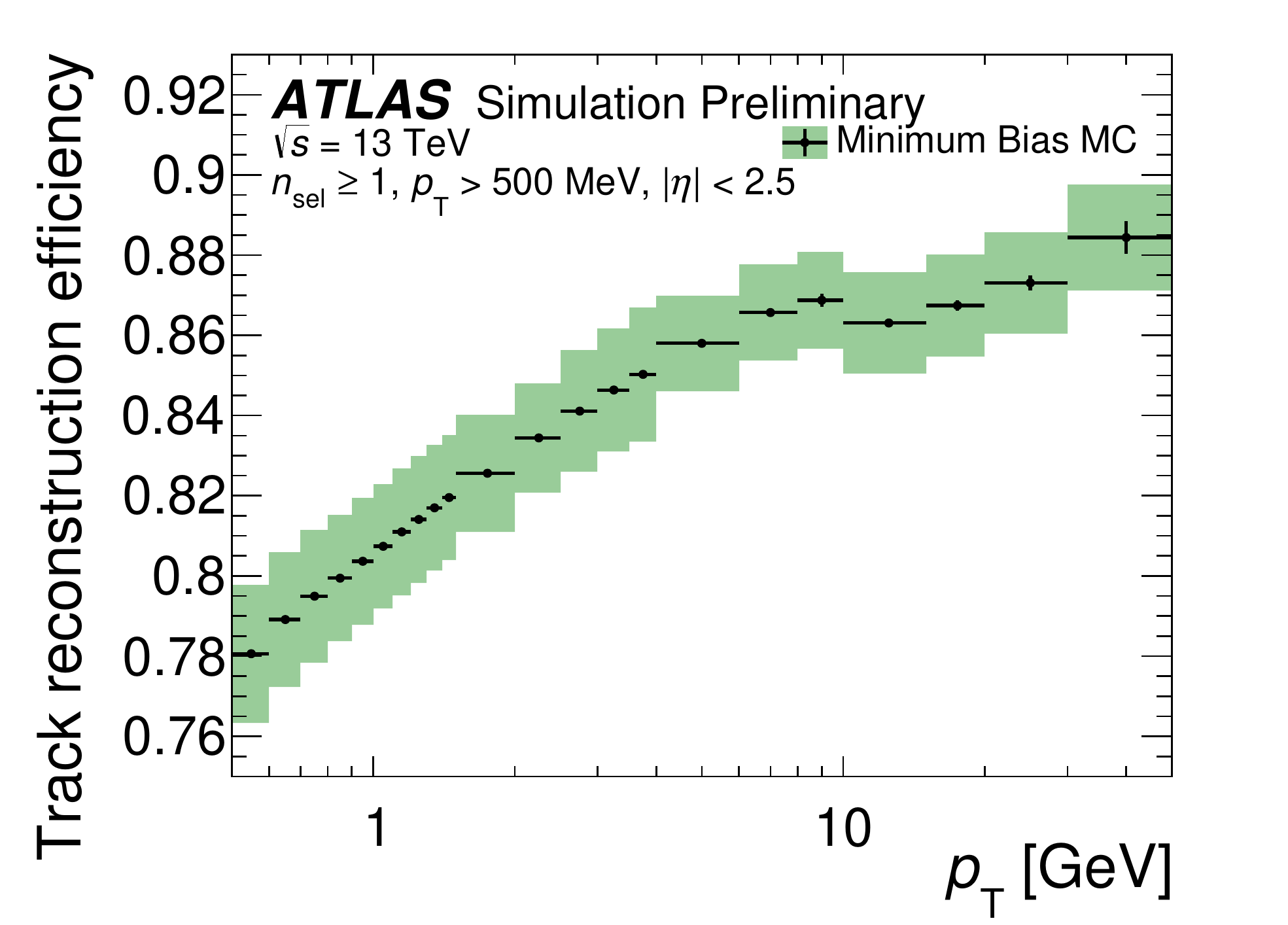}
   %             \label{fig:eff-trackpt}
       % }
        \end{center}
        \caption[]{The track reconstruction efficiency as a function of 
(left) pseudorapidity, $\eta$, and (right) transverse momentum, \pta.
The statistical errors are shown as black lines, the total errors 
as green shaded areas~\cite{ATLAS-CONF-2015-028}.}
  \label{fig:efftrk}
\end{figure}

\begin{figure}
        \begin{center}
       % \subfigure[]{
                \includegraphics[width=0.48\textwidth]{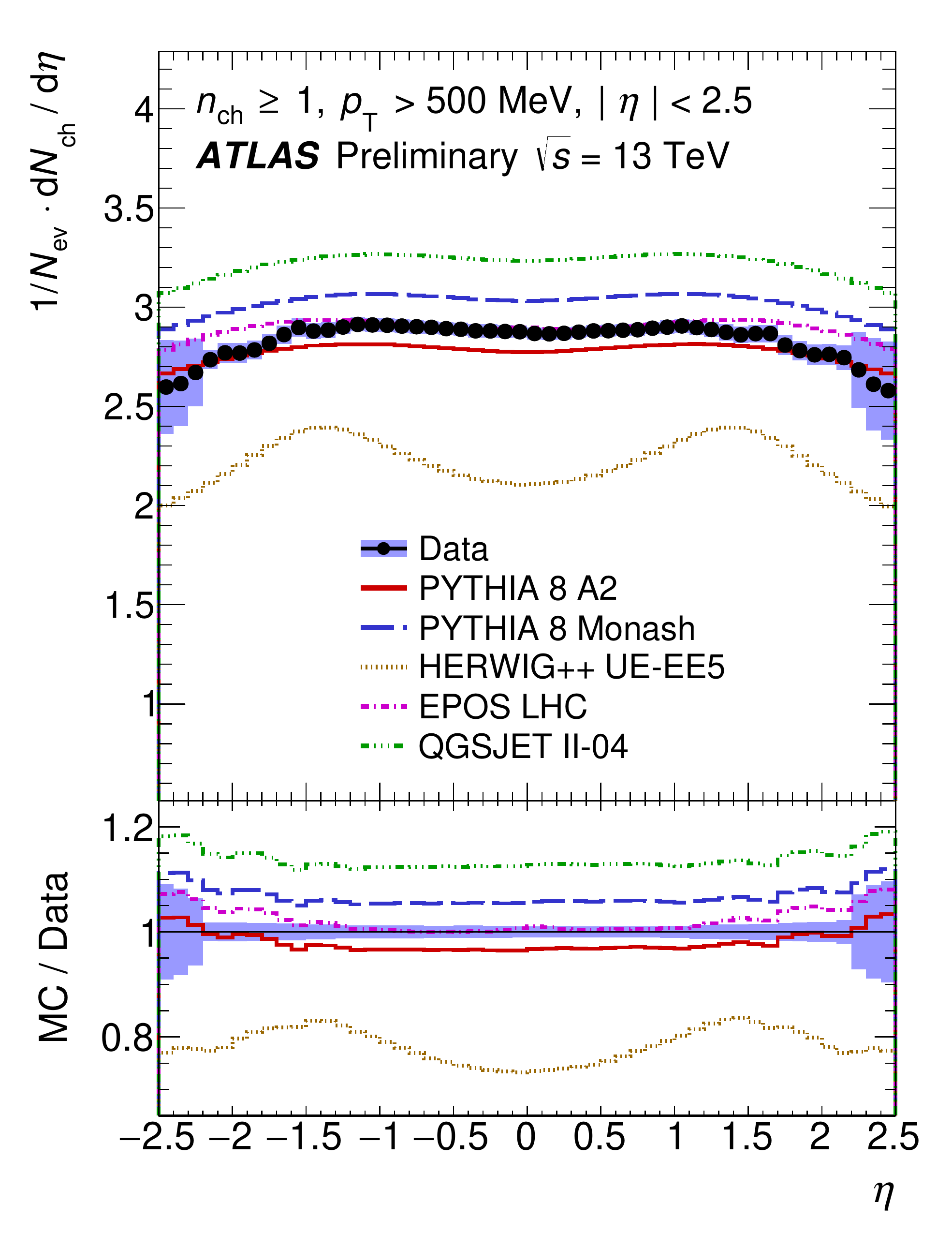}
     %           \label{fig:multeta_a}
       % }
       % \subfigure[]{
                \includegraphics[width=0.48\textwidth]{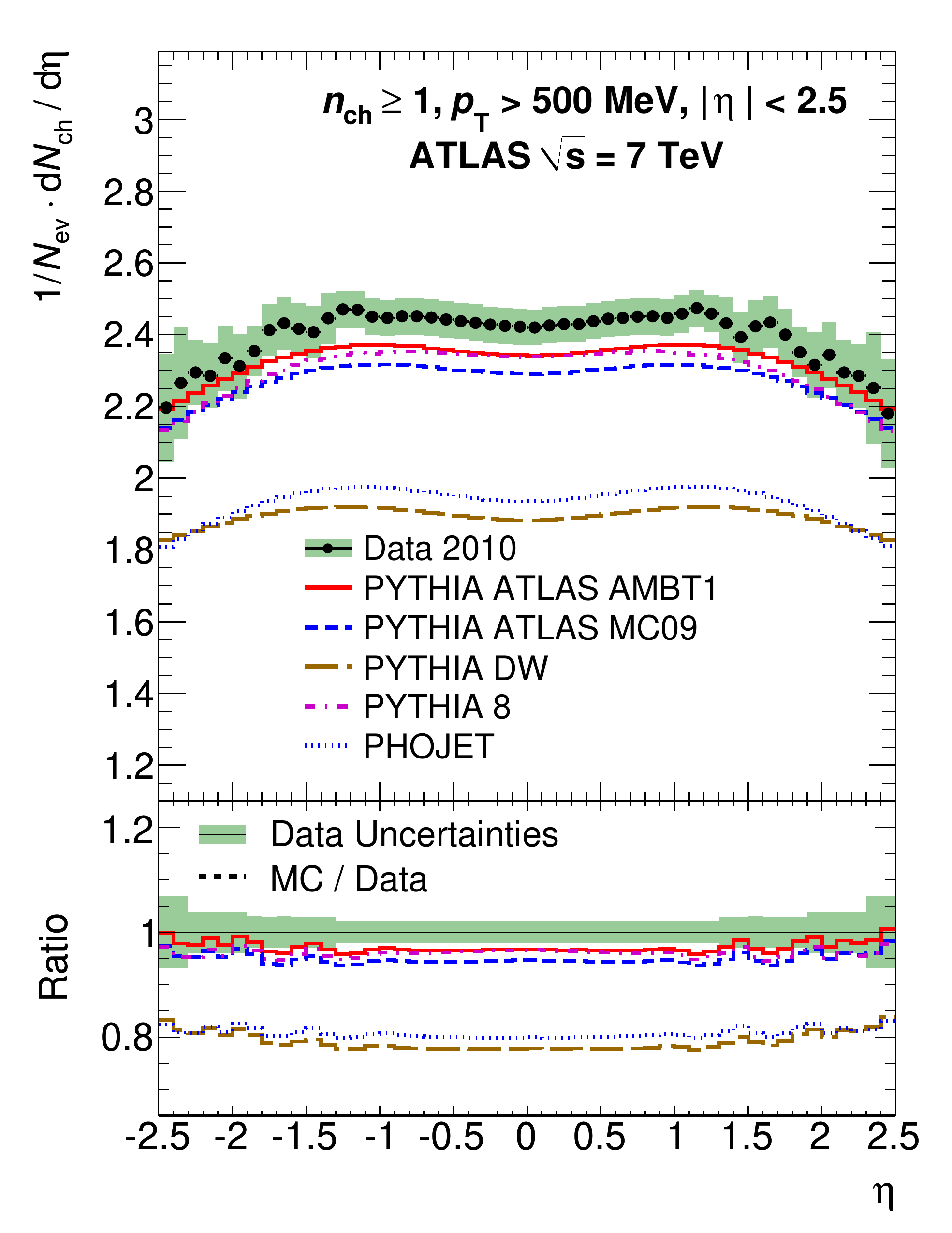}
     %           \label{fig:multeta_b}
       % }
        \end{center}
 \caption[]{Charged-particle multiplicity as a function of the
pseudorapidity for events with  $\nch \geq 1$, $\pta > 500$~MeV and $|\eta| < 2.5$ at
\mbox{(left) $\sqrt{s}=13$~TeV~\cite{ATLAS-CONF-2015-028}} and 
(right) $\sqrt{s}=7$~TeV~\cite{STDM-2010-06}. 
The dots represent the data and the curves the predictions from different MC models. The vertical bars represent the statistical uncertainties,
while the shaded areas show statistical and systematic uncertainties added in quadrature.
The bottom inserts show the ratio of the MC over the data. The values
of the ratio histograms refer to the bin centroids.}
  \label{fig:multeta}
\end{figure}

%-------------------------------------------------------------------------------
\section{Charged-particle distributions}
\label{sec:mult}
%-------------------------------------------------------------------------------

In the first presented analysis the properties of the charged particles
produced in the minimum bias $pp$ collisions at $\sqrt{s}=13$~TeV were 
studied~\cite{ATLAS-CONF-2015-028}. The reconstructed tracks are required to 
satisfy the following criteria:
\begin{itemize}
 \item $\pta>0.5$~GeV and $|\eta|<2.5$,
 \item at least 1 pixel hit with the additional requirement of an innermost pixel detector hit if expected
(if not expected then next to innermost hit is required, if expected),
 \item at least 6  hits in the SCT detector,
 \item  $\dzerobs<1.5$ mm (the transverse impact parameter, \dzerobs,
 is calculated with respect to the average beam position),
  \item $|\Delta z_{0}\cdot \sin \theta|<1.5$~mm ($\Delta z_{0}$ is
the difference between the longitudinal position of the track along the beam
line at the point where \dzerobs is measured and the longitudinal 
position of the primary vertex),
 \item $\chi^{2}$ probability $> 0.01$ for tracks with $\pta>10$~GeV.
\end{itemize}

The efficiency of the track reconstruction is obtained from MC
simulations. 
It is defined as the number of reconstructed tracks
matched to charged primary particles divided by the number of 
charged primary particles. It is a function of 
both $\pta$ and $\eta$ as one can see in Fig.~\ref{fig:efftrk}.

In the calculations of the final results several corrections are 
applied. The trigger efficiency and the vertex reconstruction efficiency
are used to account for the loss of events. For distributions of $\pta$ 
and $\eta$ it is necessary to apply track reconstruction efficiency correction  
and subtract contributions from secondary particles, strange baryons and particles
from outside of the kinematic range. Finally, the multiplicity distribution
is calculated using the Bayesian unfolding.

\begin{figure}
        \begin{center}
       % \subfigure[]{
                \includegraphics[width=0.48\textwidth]{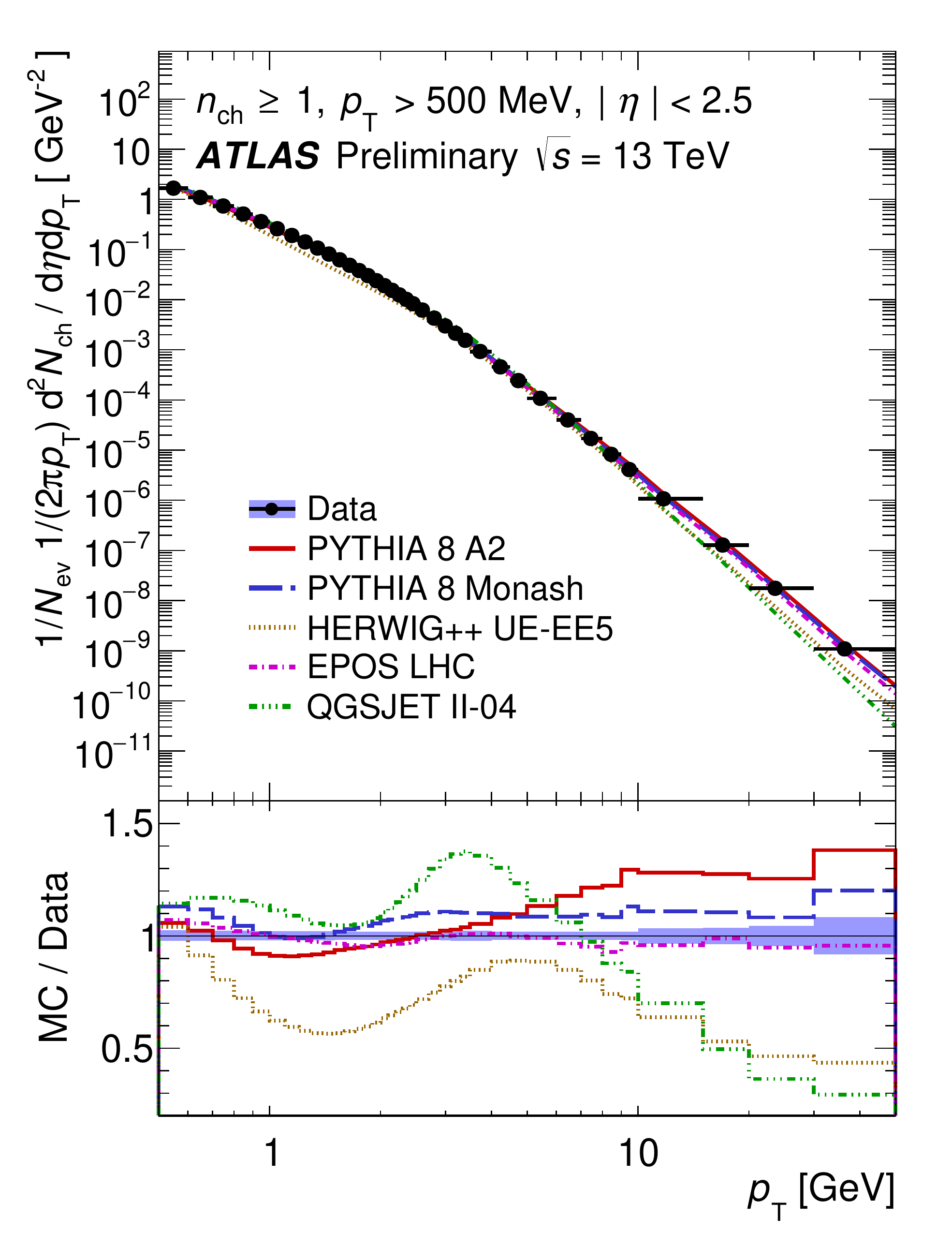}
  %              \label{fig:multpt_a}
       % }
       % \subfigure[]{
                \includegraphics[width=0.48\textwidth]{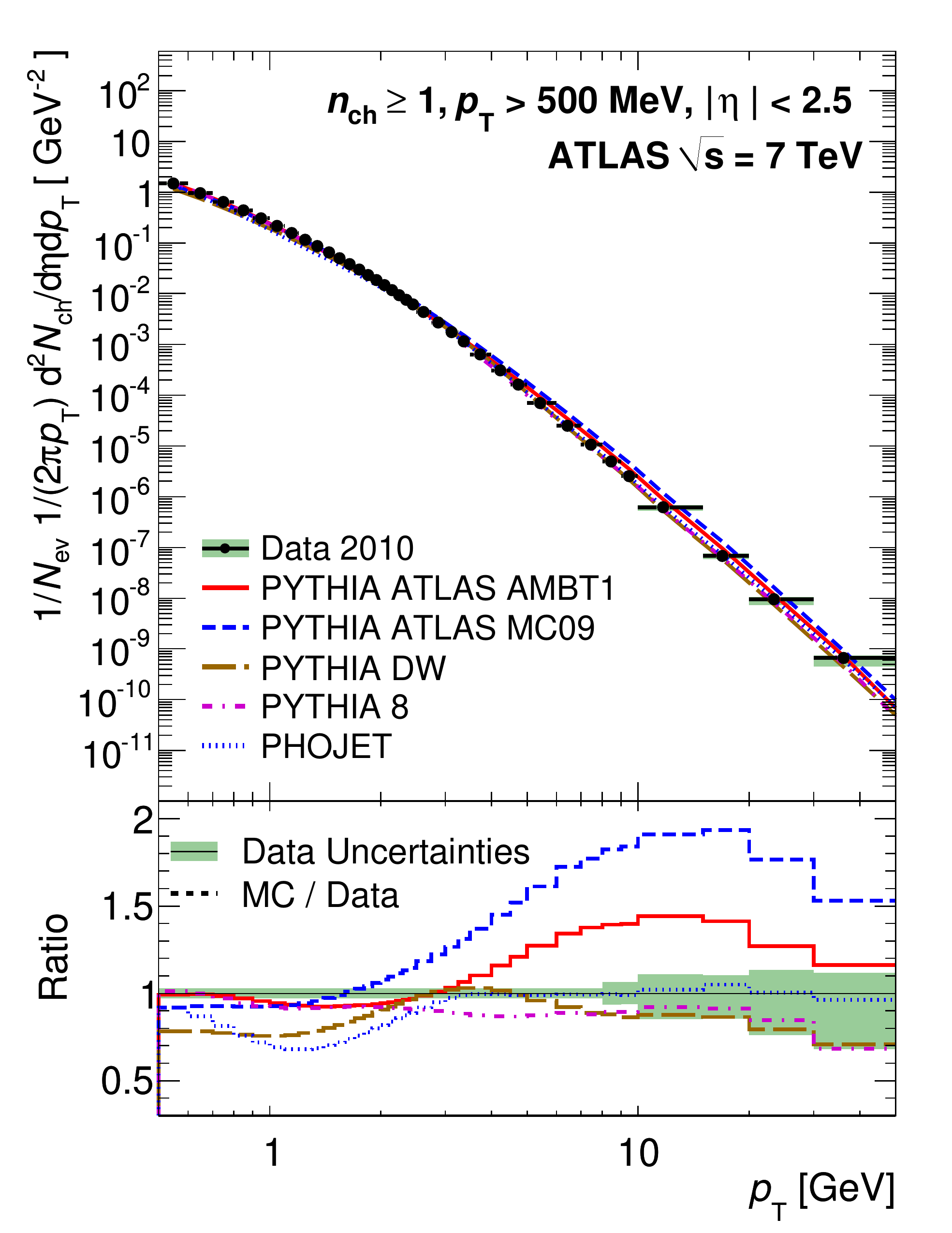}
  %              \label{fig:multpt_b}
       % }
        \end{center}
 \caption[]{Charged-particle multiplicities as a function of the
transverse momentum for events with 
 $\nch \geq 1$, $\pta > 500$~MeV and $|\eta| < 2.5$
at (left) $\sqrt{s}=13$~TeV~\cite{ATLAS-CONF-2015-028} and
(right) $\sqrt{s}=7$~TeV~\cite{STDM-2010-06}.
The meaning of the symbols is described in the caption of Fig.~\ref{fig:multeta}.}
  \label{fig:multpt}
\end{figure}

\begin{figure}
        \begin{center}
       % \subfigure[]{
                \includegraphics[width=0.48\textwidth]{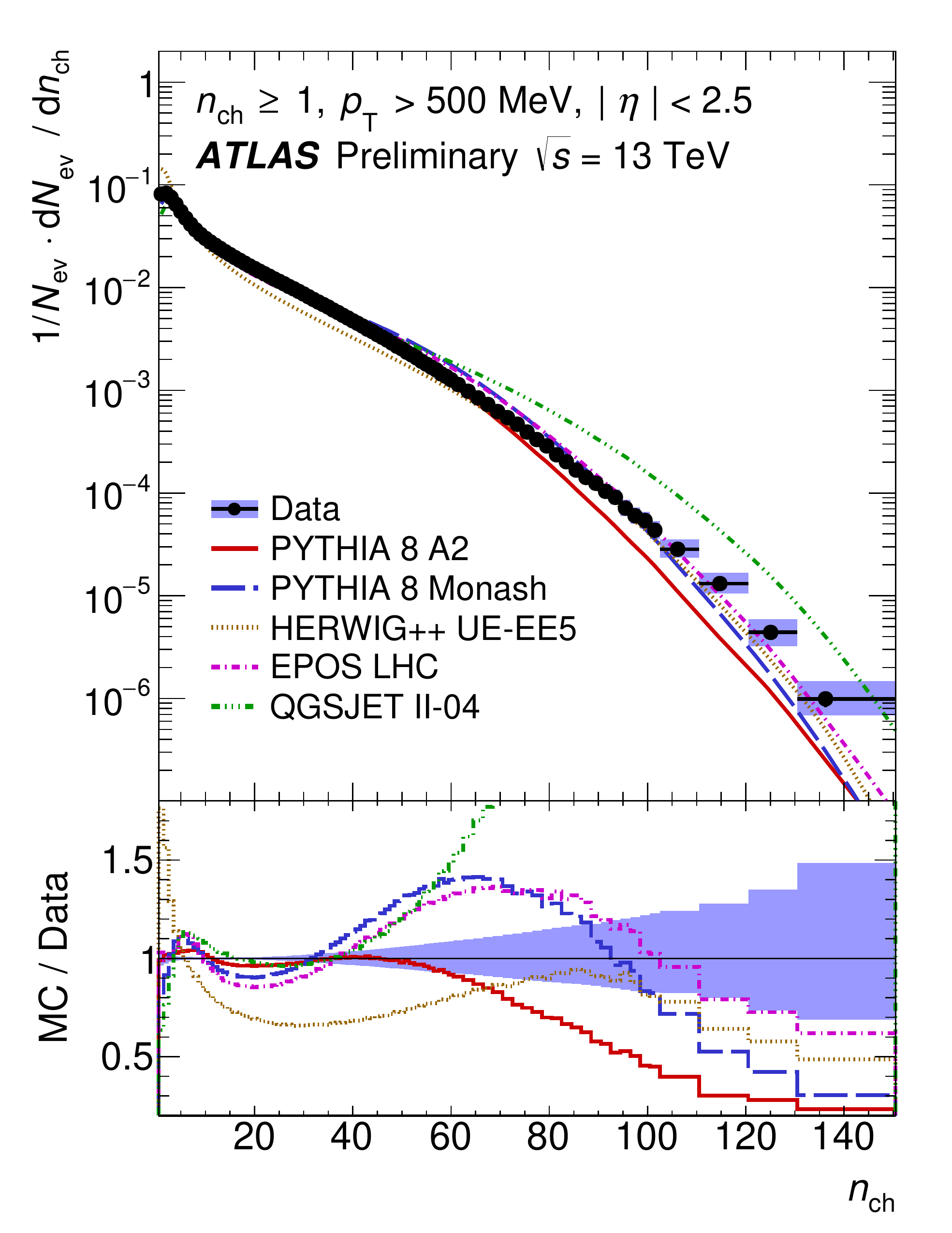}
  %              \label{fig:multnch_a}
       % }
       % \subfigure[]{
                \includegraphics[width=0.48\textwidth]{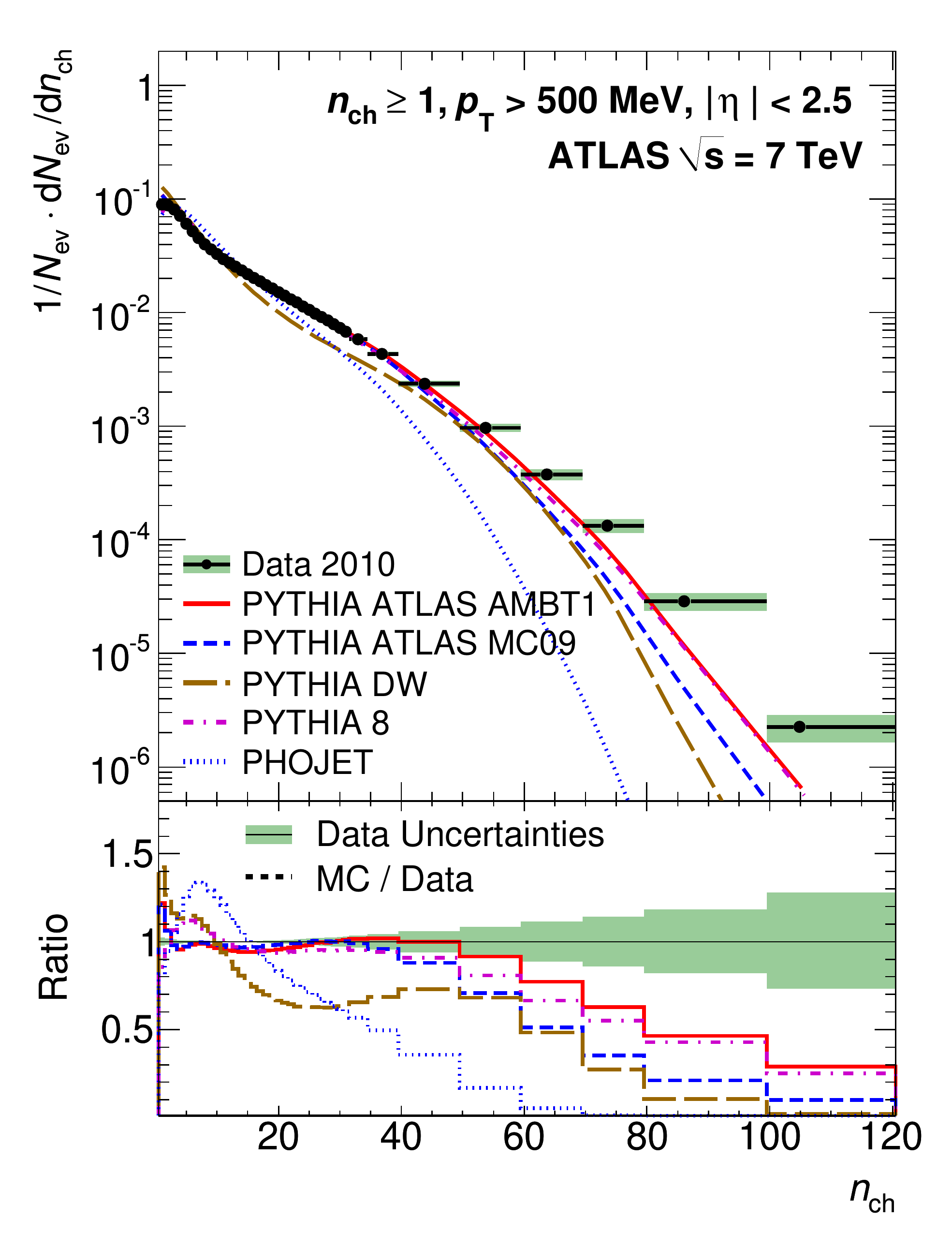}
  %              \label{fig:multnch_b}
       % }
        \end{center}
 \caption[]{The multiplicity distribution
  for events with $\nch \geq 1$, $\pta > 500$~MeV and $|\eta| < 2.5$
at \mbox{(left) $\sqrt{s}=13$~TeV~\cite{ATLAS-CONF-2015-028}} and
(right) $\sqrt{s}=7$~TeV~\cite{STDM-2010-06}.
The meaning of the symbols is described in the caption of Fig.~\ref{fig:multeta}.}
  \label{fig:multnch}
\end{figure}

In Fig.~\ref{fig:multeta} the pseudorapidity distribution,
\dndeta, is shown and compared with the MC model predictions.
Most models correctly describe the $\eta$ dependence even if the
mean multiplicity is not exactly reproduced. The one exception is 
the prediction from \hpp{}, which differs both in the height and in the
shape of this distribution. The comparison with the similar results
of the analysis at $\sqrt{s}=7$~TeV 
(Fig.~\ref{fig:multeta}~(right))~\cite{STDM-2010-06} does not indicate
any qualitative differences, only the increase of the  charged particle 
multiplicity with energy by about 15\%. The model predictions at the time when the 7~TeV
data were measured generally underpredicted 
the total charged particle multiplicity.

\begin{figure}[htb!]
\begin{center}
\includegraphics[width=0.6\textwidth]{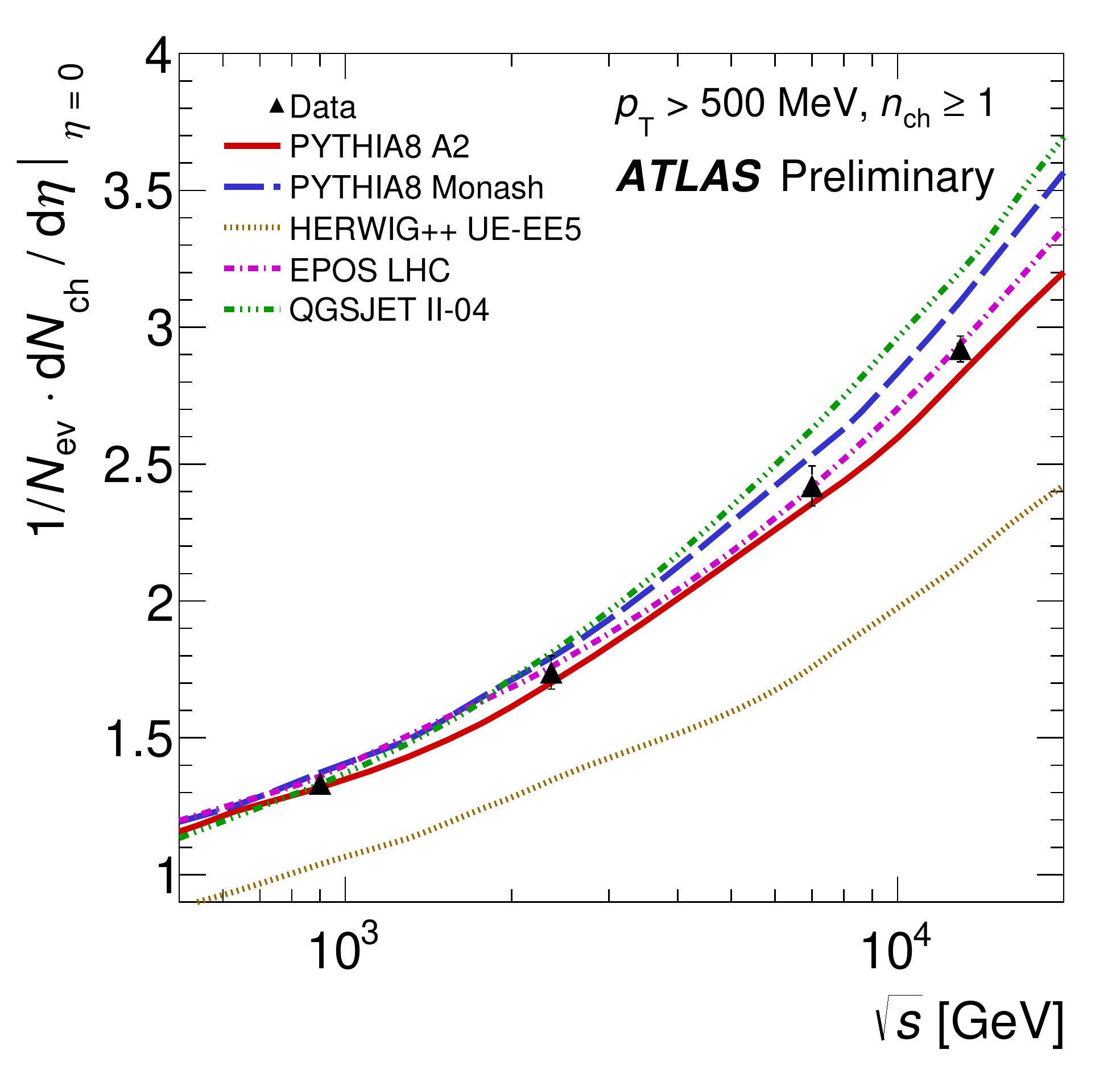}
\caption[]{The average charged-particle multiplicity per unit of rapidity 
for $\eta$ = 0 as a function of the centre-of-mass energy~\cite{ATLAS-CONF-2015-028}.
The data are compared to various particle level MC predictions.
The vertical error bars on the data represent the total uncertainty.}
\label{fig:multsqrts}
\end{center}
\end{figure}

The transverse momentum distribution, \dndpt, shown in Fig.~\ref{fig:multpt},
is measured for $\pta<50$~GeV
and spans over 9 orders of magnitude. MC models appear similar to the data
when this distribution is presented in the log scale, but
the relative differences are large, especially for \hpp{}  
and, at high $\pta$, for \qgsjet{}.
The distribution obtained for 7~TeV data has a similar shape.
The comparison with the predictions of MC models revealed in some
cases even larger differences than at 13~TeV.

The largest discrepancies between data and MC predictions
are observed in the case of multiplicity distribution, \dndeta,
shown in Fig.~\ref{fig:multnch}.
The \hpp{} is farthest from the data at very low ($\nch<5$) and at
moderate multiplicities (10-50), while for $\nch>60$ \qgsjet{} deviates
most significantly. None of the models predicts correctly this
distribution for the highest multiplicities. Similar conclusions
are valid also in the case of 7~TeV data, only the models
compared are different.

% The models considered most precisely reproduce the dependence of
% the mean transverse momentum on \nch (not shown, see~\cite{ATLAS-CONF-2015-028}), for which
% only the predictions of \qgsjet{} deviate more than 10\%, which is
% not surprising as this

The collision energy dependence of the mean number of charged particles
in the central region (calculated for $|\eta|<0.2$) is shown
in Fig.~\ref{fig:multsqrts}. The multiplicity increases faster than linearly,
and model predictions are qualitatively similar. The closest
to the data is \epos{}, while \hpp{} underpredicts multiplicity for all energies.

% ===============================================
% UE

%-------------------------------------------------------------------------------
\section{Underlying event analysis}
\label{sec:UE}
%-------------------------------------------------------------------------------

\begin{figure}
        \begin{center}
       % \subfigure[]{
                \includegraphics[width=0.49\textwidth]{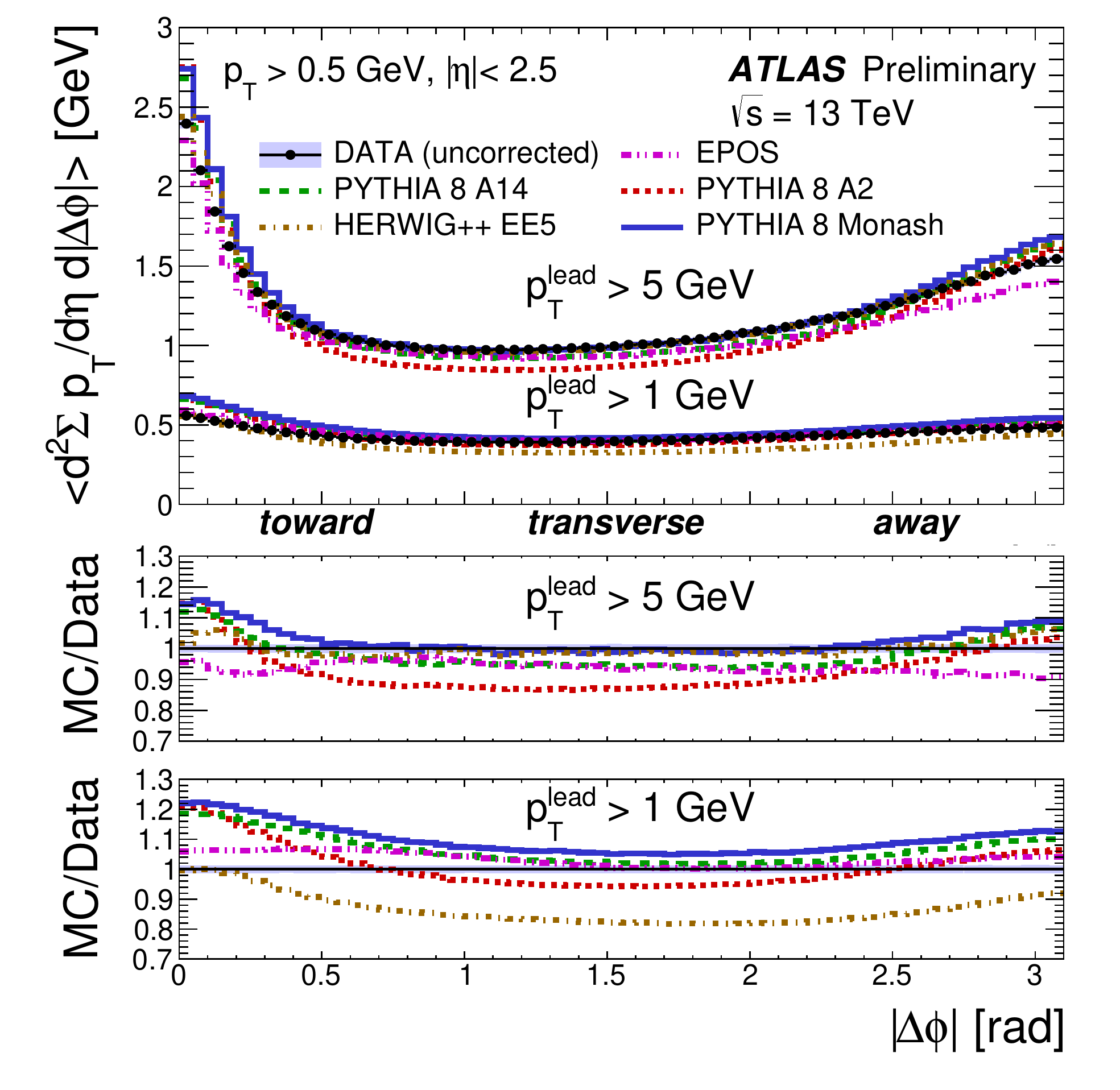}
                \label{fig:ue13_phi_a}
       % }
       % \subfigure[]{
                \includegraphics[width=0.49\textwidth]{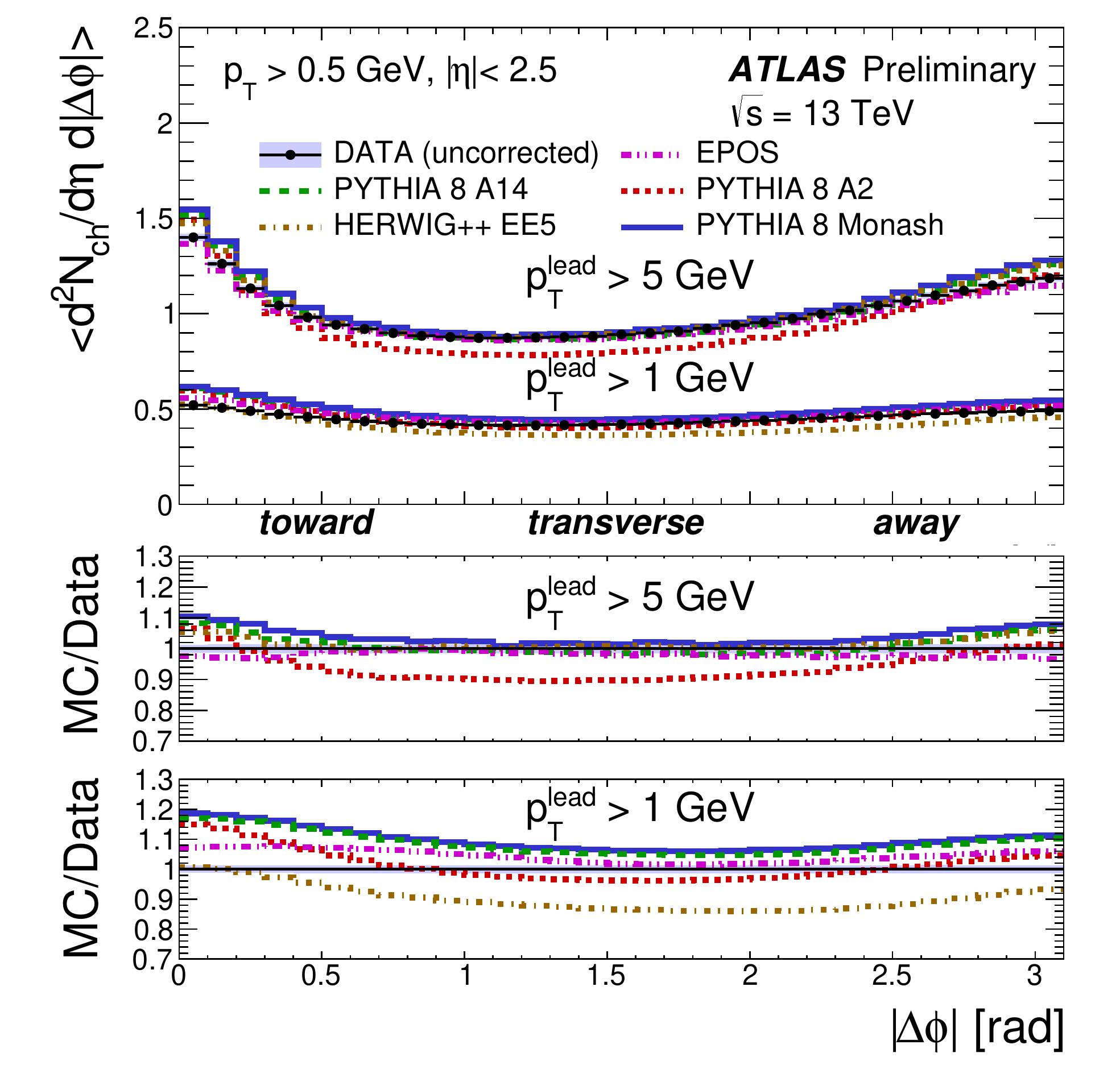}
                \label{fig:ue13_phi_b}
       % }
        \end{center}
 \caption[]{
Comparison of detector level data and MC predictions for 
the $|\Delta\phi|$ distributions of 
(left) the average scalar $\pta$ sum density of tracks and
(right)  the average track multiplicity density 
for two different minimal values of $\ptlead{}>1$~GeV 
and $\ptlead{}>5$~GeV~\cite{ATL-PHYS-PUB-2015-019}.
The leading track, defined to be at $\phi=0$, is excluded from 
the distributions. 
The bottom panels in each plot show the ratio of MC predictions to data. 
The shaded bands represent the combined statistical and systematic 
uncertainties, while the error bars show the statistical uncertainties.
}
  \label{fig:ue13_phi}
\end{figure}

\begin{figure}
        \begin{center}
       % \subfigure[]{
                \includegraphics[width=0.49\textwidth]{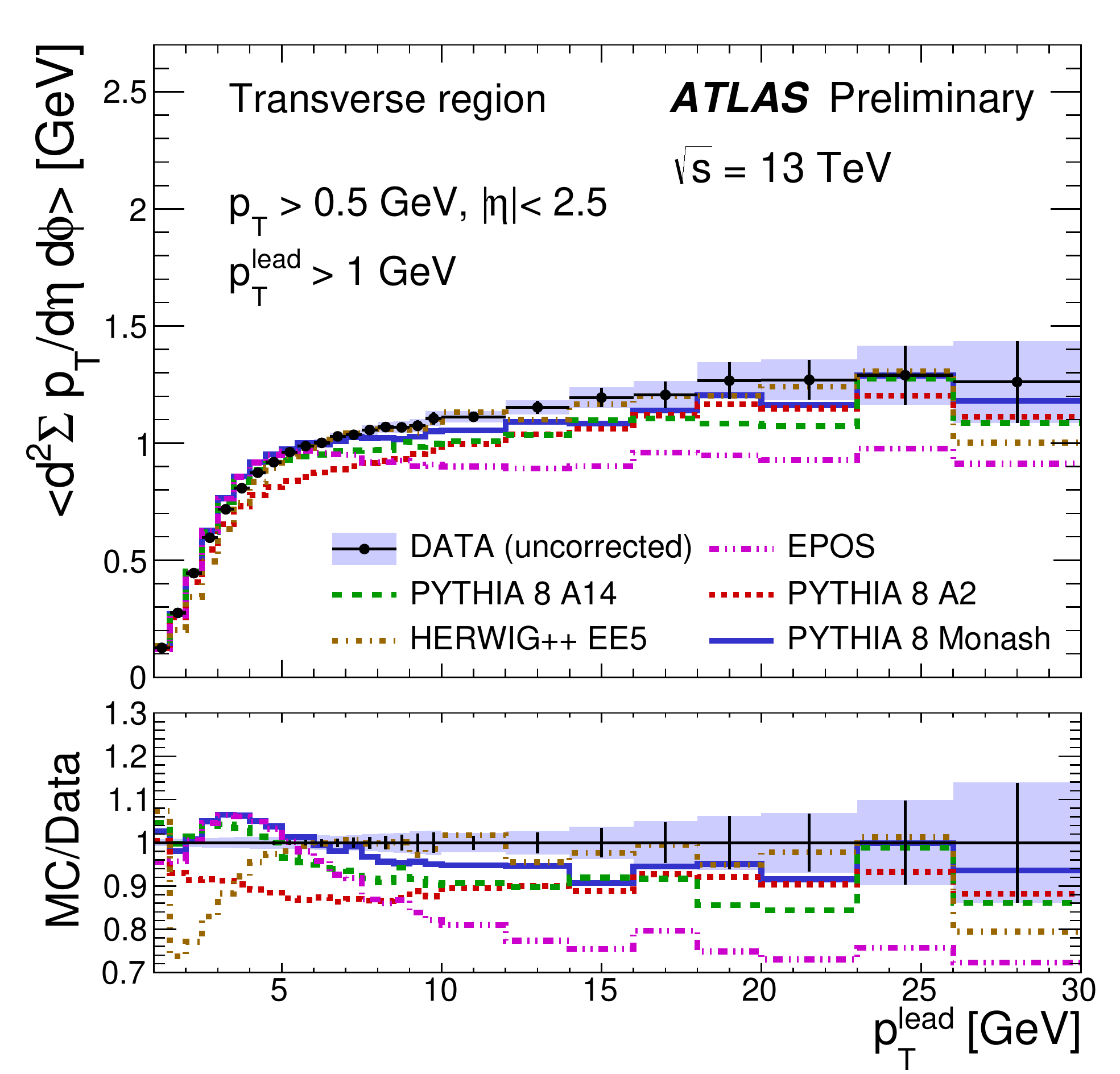}
%                \label{fig:ue13_pt_a}
       % }
       % \subfigure[]{
                \includegraphics[width=0.49\textwidth]{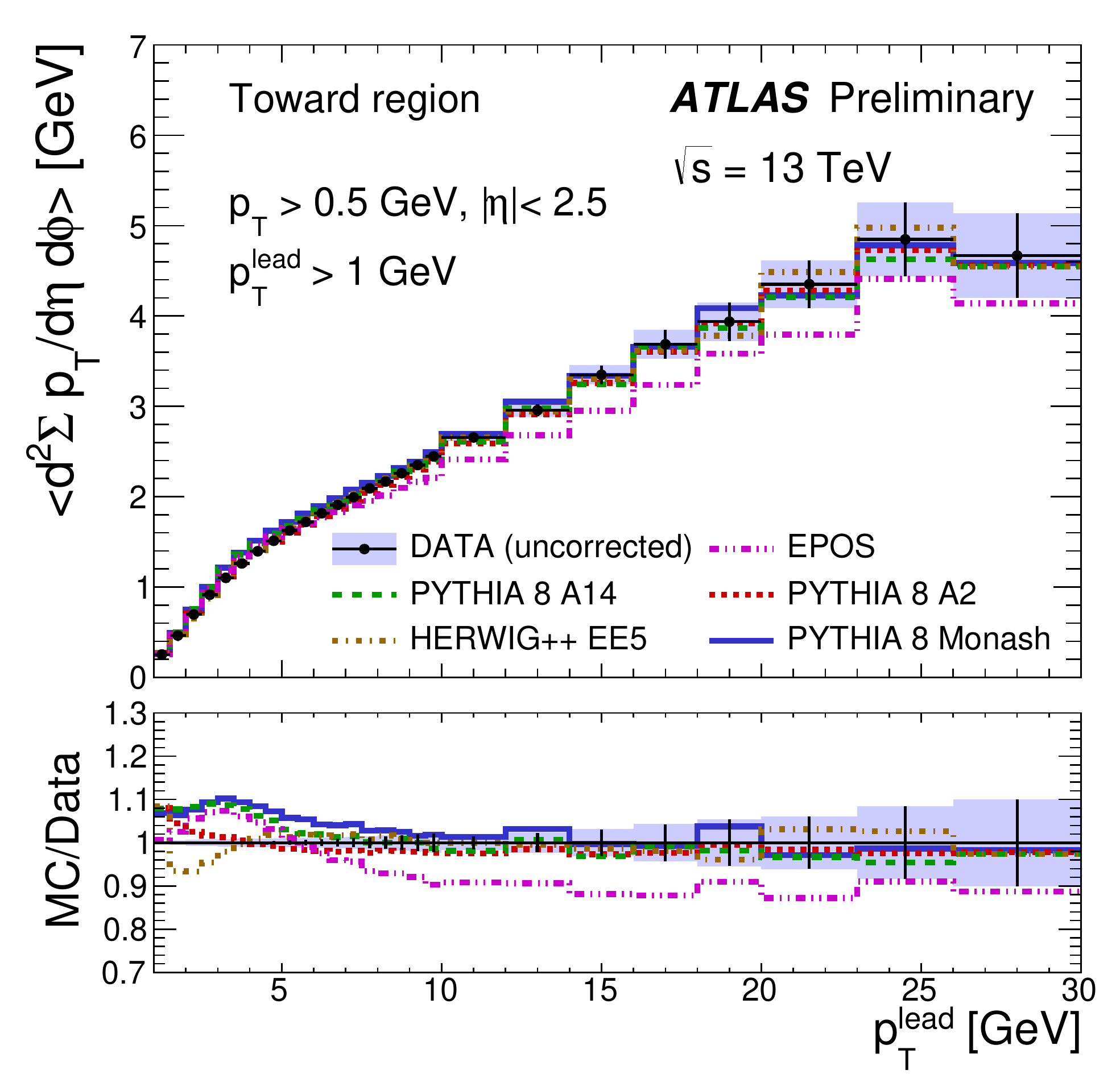}
%                \label{fig:ue13_pt_b}
       % }
        \end{center}
 \caption[]{
Comparison of detector level data and MC predictions for 
the average scalar $\pta$ sum density of tracks 
 as a function of leading track transverse momentum, \ptlead{},
in the (left) transverse and (right) toward regions~\cite{ATL-PHYS-PUB-2015-019}.
The bottom panels in each plot show the ratio of MC predictions to data. The shaded bands represent the combined statistical and systematic uncertainties, while the error bars show the statistical uncertainties.
}
  \label{fig:ue13_pt}
\end{figure}

Many studies of $pp$ interactions are devoted to the processes requiring 
a hard scattering of constituents of the colliding protons. 
Particles produced in such processes are however always accompanied 
by other particles originating from remaining partons, 
multi-parton interactions and initial or final state gluon radiation. 
Such activity not connected with the hard scattering is called 
the underlying event (UE). Even if unambiguous separation of the hard 
scattering and UE in each event is not possible, we can study 
averaged properties of UE. Usually, event properties 
in 3 separated regions are analysed, according to the value of 
$\Delta\phi = \phi-\phi_{lead}$, which is calculated 
for all tracks with respect to the azimuthal angle of a leading object, 
$\phi_{lead}$. The three regions are defined as:
\begin{itemize}
 \item $toward$ region: $|\Delta\phi|<60^{\circ}$,
 \item $transverse$ region: $60^{\circ}<|\Delta\phi|<120^{\circ}$,
 \item $away$ region: $120^{\circ}<|\Delta\phi|$.
\end{itemize}
In the studies presented here the leading object is either 
a high-$\pta$ track, a leading jet or a Z boson. 
In the first study of the 13~TeV data only leading tracks are 
considered~\cite{ATL-PHYS-PUB-2015-019} while all types of leading objects were 
analysed using 7~Tev 
data~\cite{STDM-2010-04,STDM-2010-05,STDM-2011-30,STDM-2011-31,STDM-2011-42}.

In Fig.~\ref{fig:ue13_phi} the average
scalar $\pta$ sum density of tracks and the average track 
multiplicity density as a function of $|\Delta\phi|$ 
for the 13~TeV data and the MC model predictions are shown. 
The data and event selection are the same as in the multiplicity 
analysis described in Section~\ref{sec:mult}.
Comparison is performed on the detector level results, 
without any efficiency corrections. 
Two cases with different cuts for the leading track transverse momentum,
one for $\ptlead{}>1$~GeV the other for $\ptlead{}>5$~GeV, are considered.  
For the higher  \ptlead{} cut the distributions show enhancement
in the toward and away regions, while for lower \ptlead{} cut 
they are almost flat. MC models describe the data within $\pm 20$\%.

The dependence of the average scalar $\pta$ sum density of tracks on 
the transverse momentum of the leading track is shown in Fig.~\ref{fig:ue13_pt}
for the transverse and toward regions. In the transverse region
after an initial strong increase the scalar $\pta$ sum grows weakly,
while in the toward region the increase continues up to the highest
values of \ptlead{}. MC models generally reproduce the trends 
observed in the data, but do not describe 
very well the initial increase for low  \ptlead{}. 
At larger  \ptlead{} values the largest differences are 
observed for \epos{}, which however is tuned to describe Minimum Bias events
rather than UE.

\begin{figure}
        \begin{center}
       % \subfigure[]{
                \includegraphics[width=0.49\textwidth]{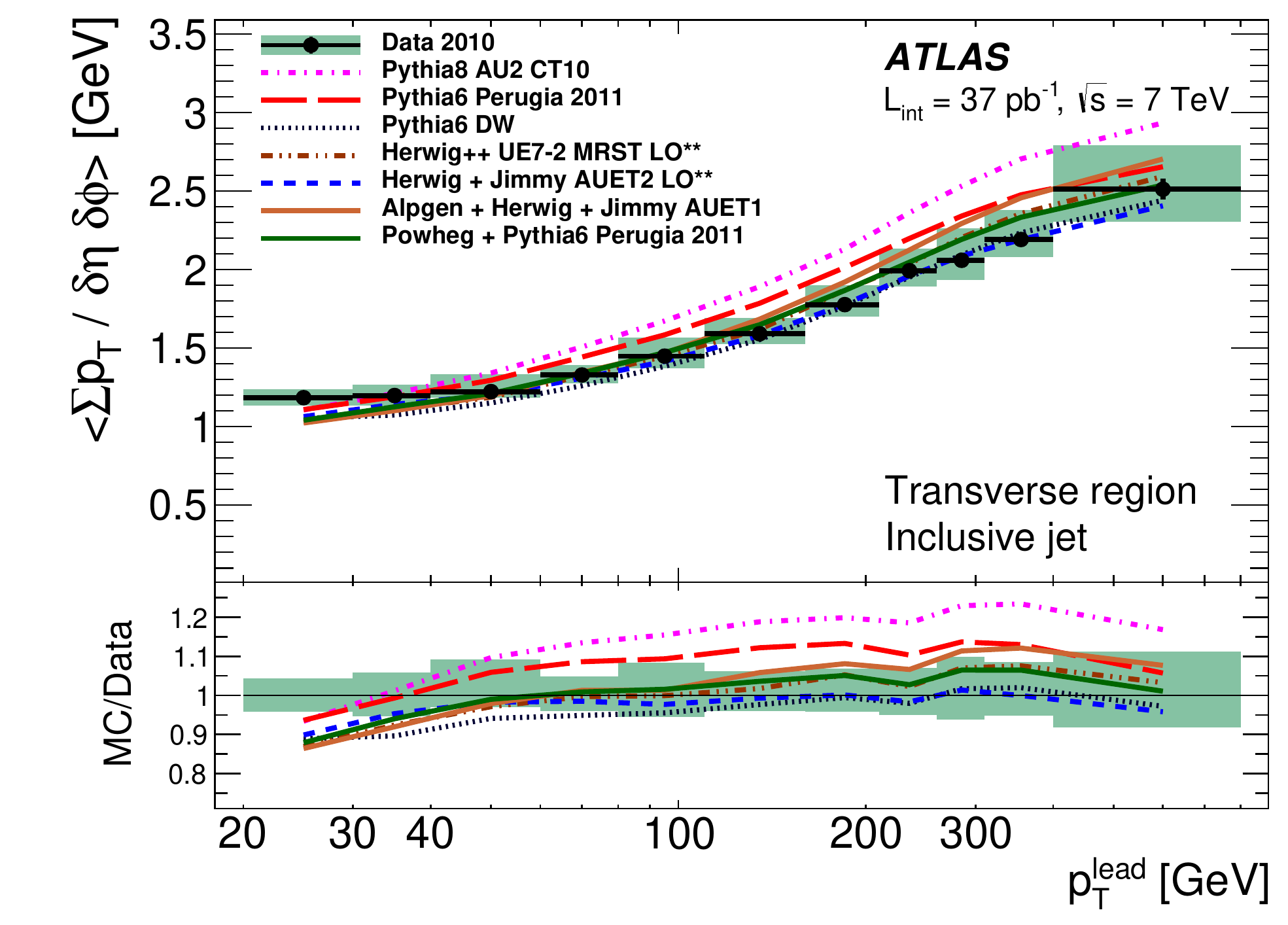}
                \label{fig:ue7jet_pt_a}
       % }
       % \subfigure[]{
                \includegraphics[width=0.49\textwidth]{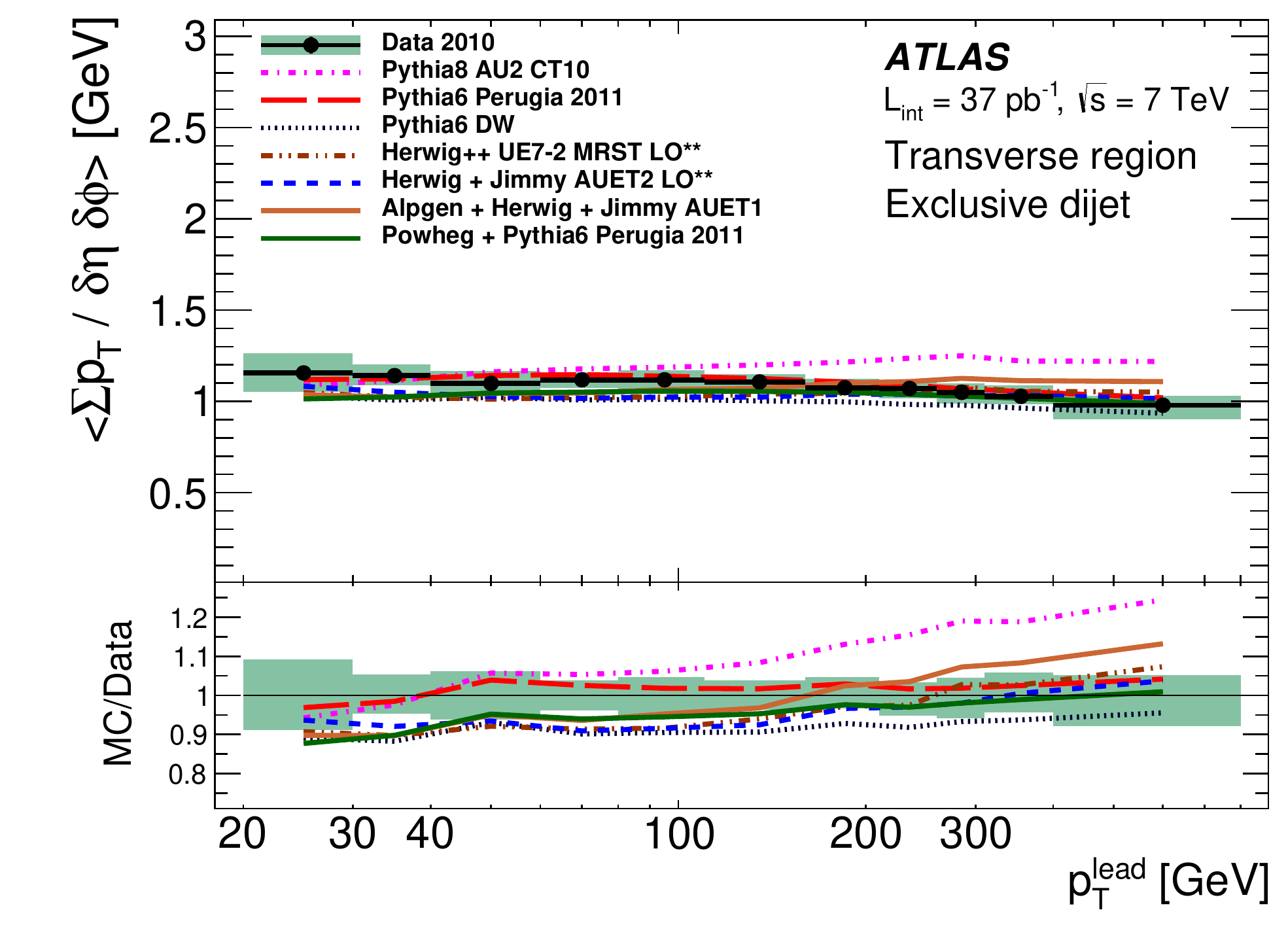}
                \label{fig:ue7jet_pt_b}
       % }
        \end{center}
 \caption[]{
The average scalar $\pta$ sum density of tracks as a function of
the leading-jet \ptlead{}
for (left) the inclusive jet and (right) the exclusive dijet
event selection~\cite{STDM-2011-31}.
% The jets are required to have pT of at least 20 GeV, and be within |y| < 2.8, whereas the charged particles have at least a pT of 0.5 GeV.
}
  \label{fig:ue7jet_pt}
\end{figure}

\begin{figure}
        \begin{center}
       % \subfigure[]{
                \includegraphics[width=0.49\textwidth]{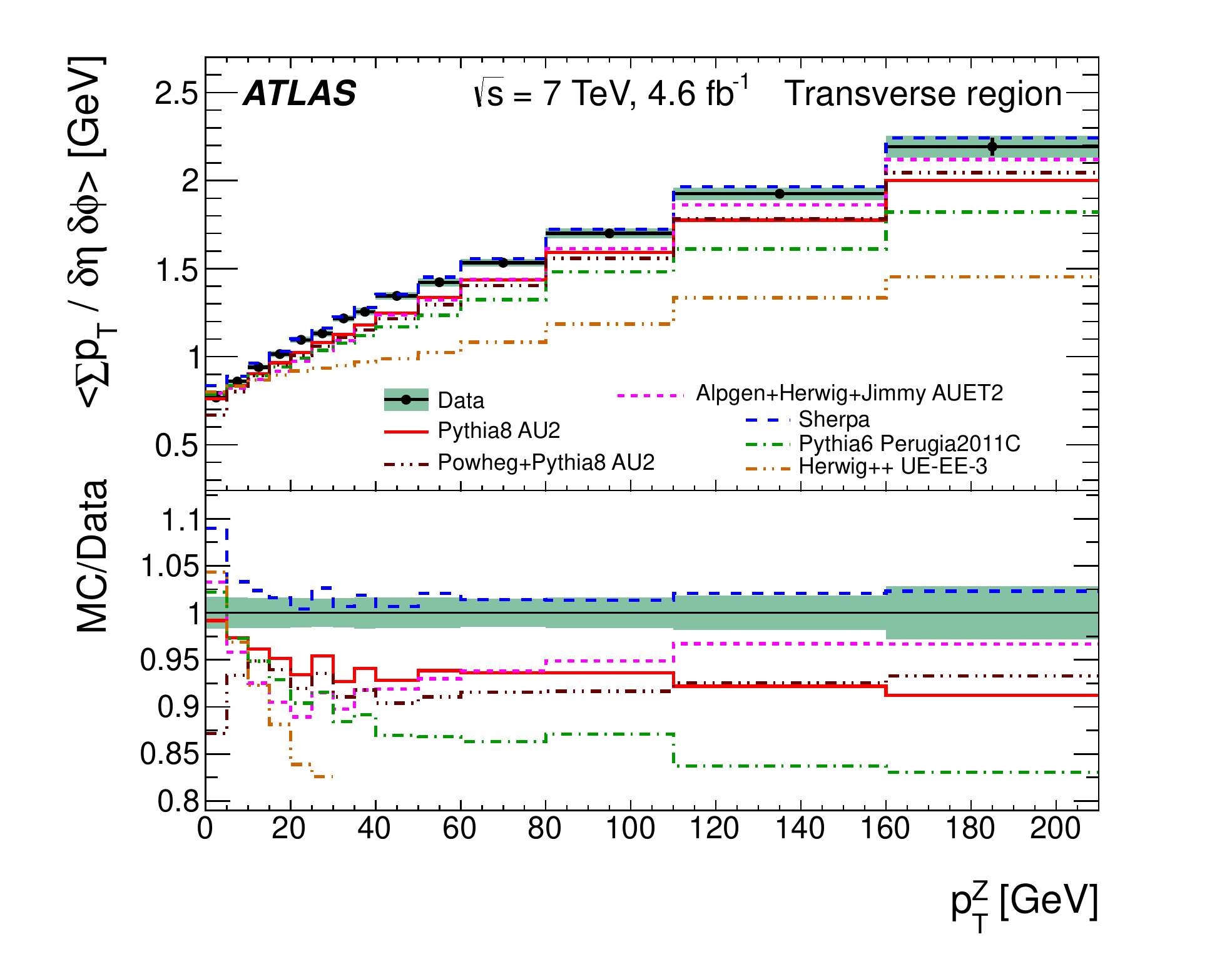}
                \label{fig:ue7z_pt_a}
       % }
       % \subfigure[]{
                \includegraphics[width=0.49\textwidth]{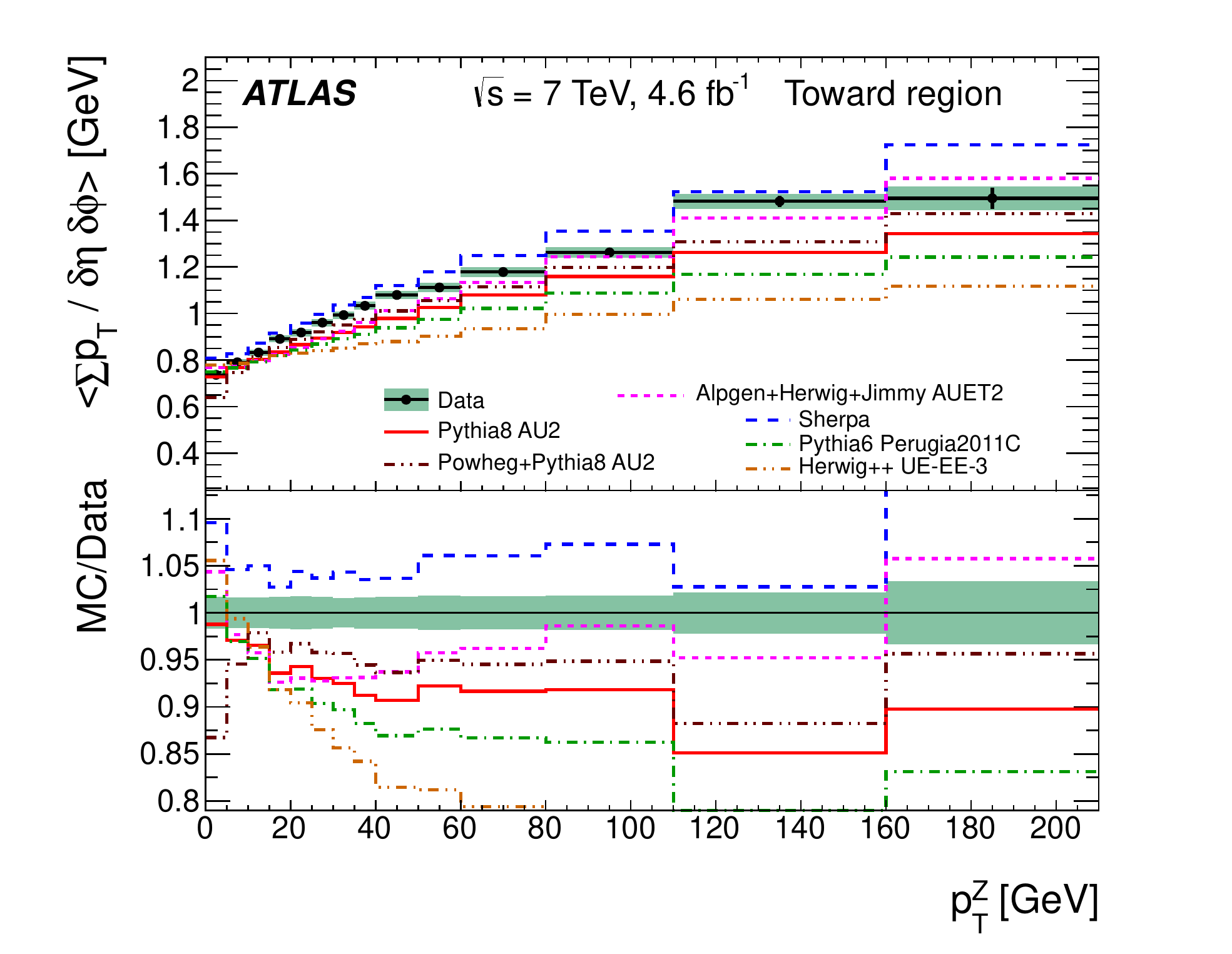}
                \label{fig:ue7z_pt_b}
       % }
        \end{center}
 \caption[]{
% Comparison of data and MC predictions for charged particle
The average scalar $\pta$ sum density as a function of Z-boson transverse 
momentum, $\pta^{Z}$, in (left) the transverse and (right) 
the toward regions~\cite{STDM-2011-42}. 
The bottom panels in each plot show the ratio of MC predictions to data. 
The shaded bands represent the combined statistical and systematic 
uncertainties, while the error bars show the statistical uncertainties 
of the model and the data.
}

  \label{fig:ue7z_pt}
\end{figure}

More exhaustive studies of underlying event were performed using 7~Tev data. Recently
jets~\cite{STDM-2011-31} and Z~bosons~\cite{STDM-2011-42} were used as the leading objects.
In Fig.~\ref{fig:ue7jet_pt} the average scalar $\pta$ sum in the transverse region
as a function of \ptlead{} in the events with a leading jet is shown.
For the inclusive sample of such events (Fig.~\ref{fig:ue7jet_pt}~(left))
we can see an increase of \sumpt{} for large values of  \ptlead{}.
This is mostly due to multi-jet events, which can be removed by requiring
exclusive dijet topology. In Fig.~\ref{fig:ue7jet_pt}~(right) the results for dijet events
are shown and no increase, but even a slight decrease, is observed.

The same \sumpt{} dependence on \ptlead{} for events with the Z-bosons as the leading
objects is shown in Fig.~\ref{fig:ue7z_pt}~(left). Here the increase with $\pta^{Z}$ is
even slightly stronger than for the events with jets, as the starting
value of \sumpt{} is ~20\% lower.
The events with a leading Z~boson can be used to study UE also in the toward region.
Here the increase of \sumpt{} with $\pta^\mathrm{Z}$ is much stronger than in the transverse region.

MC models generally describe results for leading jet better than for Z~boson. 
In the latter case predictions of the models which include production of 
additional jets coming 
from the hard scattering are closer to the data.

\begin{figure}
        \begin{center}
       % \subfigure[]{
                \includegraphics[width=0.49\textwidth]{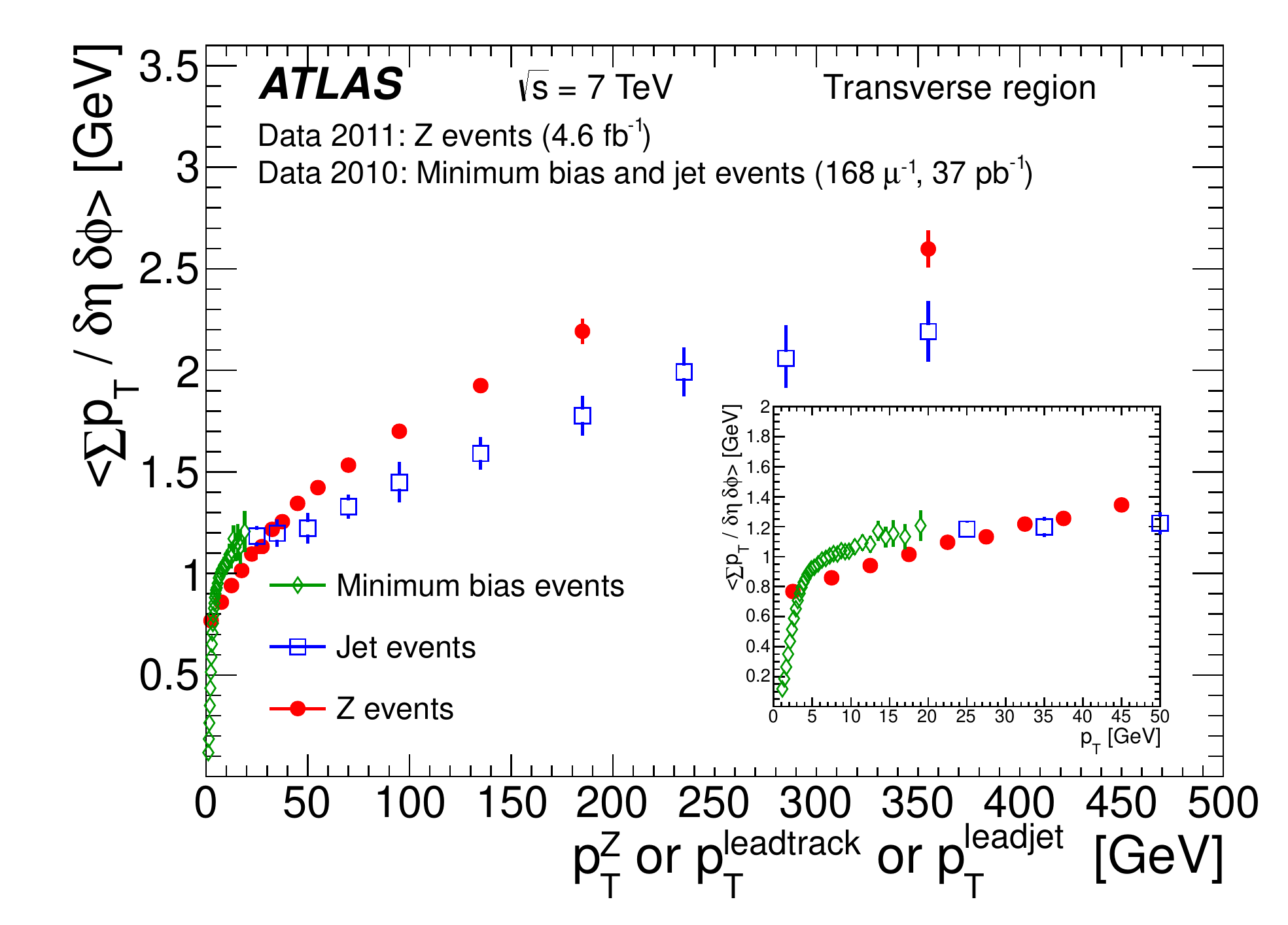}
                \label{fig:ue7z_sum_a}
       % }
       % \subfigure[]{
                \includegraphics[width=0.49\textwidth]{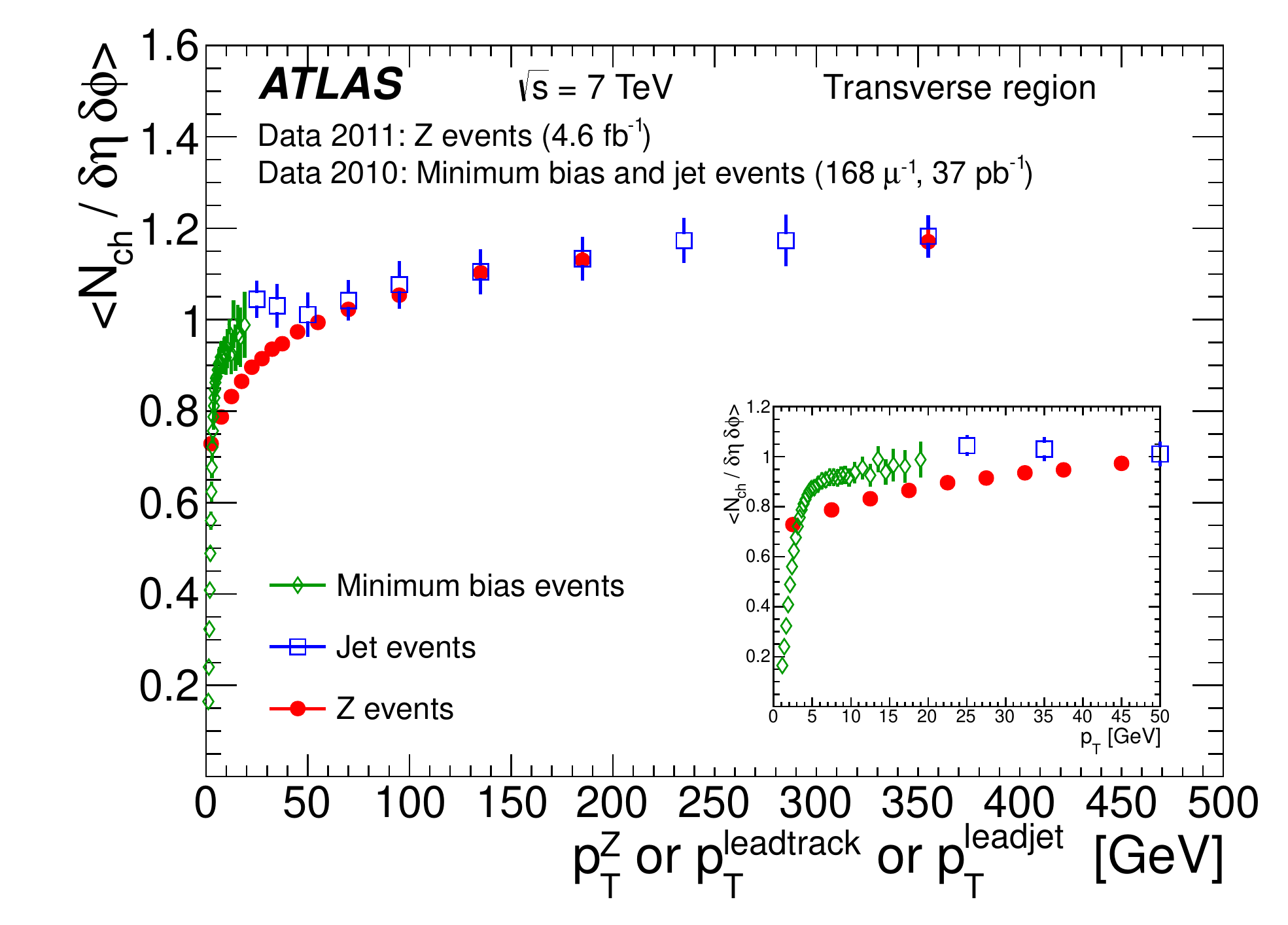}
                \label{fig:ue7z_sum_b}
       % }
        \end{center}
 \caption[]{
(left) The scalar $\pta$ sum density and 
(right) the charged particle multiplicity average values  
in the transverse region as a function of $\pta$ of 
the leading charged particle, the leading-jet and Z boson~\cite{STDM-2011-42}.  
The error bars in each case show the combined statistical and systematic uncertainties.
% The insets show the region of transition between the leading charged particle and leading jet results in more detail.
}

  \label{fig:ue7z_sum}
\end{figure}

A summary of the results of analysis of the 7~TeV data with different leading objects 
is shown in Fig.~\ref{fig:ue7z_sum}. The average values of \sumpt{} and \sumnch{} obtained 
for the leading track and the leading jet are smoothly joining at $\ptlead{}\approx 20$. 
The dependence of \sumpt{} on \ptlead{} for tracks+jets is much different than 
for Z~boson. This dependence is also different for \sumnch{}, but at  $\ptlead{}>50$~GeV 
the results for the leading jet and Z~boson are very close.

% \FloatBarrier

%-------------------------------------------------------------------------------
\section{Summary}
\label{sec:summary}
%-------------------------------------------------------------------------------

The studies of inclusive minimum bias charged-particle distributions and 
underlying event properties are among the first analyses of 
the new $pp$ data collected at $\sqrt{s}=13$~TeV.
The pseudorapidity, transverse momentum and multiplicity distributions
measured with minimal model dependence are compared to predictions
of MC models tuned to the data at lower energies. The best agreement
with the data is reached by \epos{}, a reasonable description is provided
by \py{} while \hpp{} and \qgsjet{} models are most distant from the data.
Generally, the current tunes of the models describe the data better than
the models available at the time of the first measurements of
$pp$ collisions at $\sqrt{s}=7$~TeV did.

The first UE study at $\sqrt{s}=13$~TeV on the detector level
is also aimed on the comparison with MC models.
In this case the predictions and the data agree much better, 
suggesting that the extrapolation with energy
of the multi-particle interactions in the models performs 
reasonably well. 
The general trends for the \sumpt{} dependence on \ptlead{} 
observed in the transverse region at $\sqrt{s}=13$~TeV agree with 
those found in more extensive studies performed at $\sqrt{s}=7$~TeV. 
The fast rise for $\ptlead{}<5$~GeV is continued with a moderate
slope not only for leading tracks, but also in the \ptlead{}
range available in the study of leading jets.
Similar increase of UE activity is present also in the events with
leading Z~boson, but it has a different shape. 
In contrast, for events with leading jets with a clear dijet signature
the average scalar $\pta$ sum density of tracks slightly decreases
with \ptlead{}. This means that the increase of UE activity with 
growing \ptlead{} in the transverse region can be connected
with the presence of additional low-\pta{} jets.

%-------------------------------------------------------------------------------
\section*{Acknowledgements}
%-------------------------------------------------------------------------------

%\input{acknowledgements/Acknowledgements}

This work was supported in part by the National Science Center grant 
DEC-2013/08/M/ST2/00320 and by PL-Grid Infrastructure.

%-------------------------------------------------------------------------------
\section*{References}

% \bibliography{lowx}{}

\printbibliography

%\begin{thebibliography}{99}
%\bibitem{Basar} G. Basar and D. Teaney, arXiv:1312.6770 [nucl-th].
%\end{thebibliography}

\end{document}